\documentstyle[prl,aps,multicol]{revtex}

\newcommand{\lsim}{\mathrel{\mathop{\kern 0pt \rlap
  {\raise.2ex\hbox{$<$}}}
  \lower.9ex\hbox{\kern-.190em $\sim$}}}
\newcommand{\gsim}{\mathrel{\mathop{\kern 0pt \rlap
  {\raise.2ex\hbox{$>$}}}
  \lower.9ex\hbox{\kern-.190em $\sim$}}}

\title{A test of Local Realism with entangled kaon pairs and without inequalities}

\author{Albert Bramon$^{1}$ and Gianni Garbarino$^{2}$}

\address{$^1$Grup de F{\'\i}sica Te\`orica,
Universitat Aut\`onoma de Barcelona, E--08193 Bellaterra, Spain}
\address{$^2$Departament d'Estructura i Constituents de la Mat\`{e}ria,
Universitat de Barcelona, E--08028 Barcelona, Spain} \date{\today}

\begin{document}
\draft
\maketitle

\begin{abstract}
We propose the use of entangled pairs of neutral kaons, considered as a promising tool
to close the well known loopholes affecting generic Bell's inequality tests, 
in a specific Hardy--type experiment. Hardy's contradiction without inequalities
between Local Realism and Quantum Mechanics can be translated into a feasible 
experiment by requiring ideal detection
efficiencies for {\it only one} of the observables to be alternatively measured.
Neutral kaons are near to fulfil this requirement and therefore to close the
efficiency loophole.

\end{abstract}

\pacs{PACS numbers: 03.65.-w, 03.65.Ud, 14.40.Aq} 

\begin{multicols}{2}



Bell's theorem \cite{bell} proved the incompatibility between Quantum Mechanics (QM)
and Local Realism (LR), and opened the attractive possibility to solve the 
famous Einstein--Bohr debate from a purely experimental point of view.
The various forms of Bell's inequalities \cite{bell,chsh,wigner,clh},  
which are strict consequences of LR but can be violated by QM,
have been the usual tool for such an experimental discrimination. 
This requires the use of entangled systems, such as 
the singlet state of two photons or two spin--half particles, 
as first considered by Bohm, or the formally equivalent two--kaon state:
\begin{equation}
\label{zero}
\Phi_0  = 
\left[{K}^0\bar{K}^0 - \bar{K}^0 {K}^0\right]/\sqrt{2},   
\end{equation}
also discussed in other recent analyses. 
{\it Maximally} entangled bipartite states like these show \cite{Kar95} the maximum
QM violation of Bell's inequalities. A lot of experiments violating
Bell's inequalities have been carried out. Unfortunately,
none of them has been loophole--free \cite{emilio}: the so--called {\it locality} and 
{\it detection} loophole affected these Bell--type tests. Important steps forward have been
done very recently: the experiments with entangled photons of Refs.~\cite{We98,Gi99} closed 
the locality loophole, while, by employing beryllium ions \cite{Ro01},
it has been possible to close the efficiency loophole. However, a test closing 
simultaneously both loopholes is lacking.

In 1992 Hardy \cite{Ha93} proved Bell's theorem without using inequalities for
any {\it non--maximally} entangled state composed of two two--level subsystems.
Such a proof allows, at least in principle, for a clear--cut  
discrimination between LR and QM for a fraction ($\lsim 9$\%),  
usually called Hardy fraction, of the single experimental runs.
The independent demonstration of Bell's theorem without inequalities provided by 
Greenberger {\it et al.} \cite{ghz} applies
to every single experimental run, {\it i.e.}, it is an `all {\it vs} nothing' proof.
Unfortunately, it requires entangled states consisting of three or more two--level 
subsystems, which are difficult to produce and control, whereas the
bipartite systems of Hardy's proof offer an easier use.
The only problem is that, being Hardy's proof 
related to a certain lack of symmetry of the state, it can not work for familiar maximally
entangled states like (\ref{zero}). Although this feature complicates the issue, Hardy's
treatment has been discussed and generalized \cite{boschi,Wu-Kar-Ca}. 

A few experiments with polarization--entangled photon pairs \cite{boschi,To95,giuseppe} 
tested LR {\it vs} QM by means of Bell's inequalities derived from Hardy's argument. 
Being these Hardy--tests affected by the same loopholes previously mentioned,
they were not conclusive, {\it i.e.}, they could not refute all versions of LR. 
In addition, a new and specific difficulty, 
the need to perform a `post--selection' of events, affects these experiments. 
It comes from the fact that true non--maximally entangled states are not easily 
produced. One thus starts with a factorizable state and,  
by a selective and {\it a posteriori} choice of the events to be considered
(the rest are discarded), one attempts to reproduce the partial
entanglement required by Hardy's argument. 
True non--maximally entangled states have been
produced very recently by using a spontaneous--down--conversion photon source
\cite{Eb99}. They have been used for a measurement
of the Hardy fraction, confirming QM. 

The unsatisfactory situation due to the previous loopholes could be improved by using 
entangled neutral kaon pairs. Such pairs are copiously produced in 
$\phi$--resonance decays into state (\ref{zero}) \cite{handbook} as well as 
in $p\bar{p}$ annihilation processes \cite{CPLEAR}.
The two kaons then fly apart from each other at relativistic velocities and 
easily fulfil the condition of space--like separation.  
Moreover, kaons as well as their decay products are strongly interacting particles,
thus allowing for high detection efficiencies \cite{sell}. 
Compared to photons (ions), neutral kaons seem thus to  
offer a more promising situation to close the efficiency (locality) loopholes. 
For these reasons, several papers on Bell's inequalities for the
$K^0\bar{K}^0$ system have appeared in the last years  
\cite{sell,ghi,eber,dome,uchi,bf,bn,BH,gigo,dg,Ge01,AlGi,long}. 
The fact that kaons are massive objects 
quite different from the massless photons usually considered, adds 
further interest to these analyses. However, two specific problems appear when dealing with
neutral kaons. The first comes from the interplay between strangeness
oscillations and weak decays, which makes very difficult to deduce Bell's inequalities
violated by QM. The other problem is that, contrary to
photons, whose polarization can be measured along any chosen direction, the choice in the
kaon case reduces to measure either its lifetime or its strangeness 
\cite{ghi,eber,bn,AlGi}. Nevertheless, a few
versions of QM violated and experimentally testable Bell's inequalities have been proposed for
maximally \cite{eber,bn,gigo,dg} and non-maximally \cite{AlGi,long} entangled kaon pairs.
 
The aim of this Letter is to explore the possibility to discriminate between LR and QM by
applying Hardy's proof to entangled kaons. There are two good reasons for doing this:
1) the fact that genuine QM measurements for kaons are only of
two types is not a drawback for Hardy's tests, since the latter require {\it two}
distinct measurement possibilities on each kaon; 2) kaon pairs 
produced in $p\bar{p}$ annihilation processes \cite{CPLEAR} 
(in $\phi$ decays \cite{handbook}) already appear in (can be converted, by means 
of a kaon regenerator, into) non-maximally
entangled states and are thus unaffected by post--selection problems.  
As we will see, clear progress is then achieved in closing all the mentioned loopholes.  

Let us start by considering the following non--maximally entangled state:
\begin{equation}
\label{stateN}
\Phi = \frac{ K_S K_L  - K_L K_S + R\, K_L K_L + R^\prime\, K_S K_S}
{\sqrt{ 2 + |R|^2+|R^\prime|^2}},
\end{equation}
which was originally discussed in Ref.~\cite{AlGi} and where:
\[
R =  -r\, \exp\left\{\left[-i\Delta m +(\Gamma_S - \Gamma_L)/2\right]T\right\} , \hspace{1mm}
R^\prime=-r^2/R  , \nonumber
\]
$\Delta m \equiv m_{L} - m_{S}$ is the difference between the $K_L$ and $K_S$ masses, 
while $\Gamma_S$ and $\Gamma_L$ are their respective decay widths.

At a $\phi$-factory, state $\Phi$ can be prepared in the following way \cite{AlGi}.  
$\phi$ decays produce the antisymmetric state 
(\ref{zero}) which, ignoring small $CP$ violation effects
($|\epsilon| <<1$), can also be written as 
$\phi_A = (K_S K_L - K_L K_S) /\sqrt 2$. A thin (few mm's)
neutral kaon regenerator placed along the right beam, close
to the pair creation point, converts state $\phi_A$ into 
$\phi_r \propto K_S K_L - K_L K_S +r K_S K_S -r K_L K_L$,
$r$ being the regeneration parameter. 
Values of $r$ are known to be rather small
[typically, $|r|= (1\div 5)\cdot 10^{-3}$ for 1 mm of material and 
kaon momenta below 1 GeV].
The state of Eq.~(\ref{stateN}) is then obtained from the unitary evolution 
of $\phi_r$ in free space up to a proper time $T$, after normalizing to undecayed pairs. 
To this aim, kaon pairs showing the decay of one (or both) member(s) before
$T$ have to be detected and excluded. Since this occurs prior to any measurement
employed in Hardy's test, ours is a `pre--selection' (as opposed to `post--selection') procedure. 

In experiments on $p\bar{p}$ annihilation at rest, 
the  state preparation is slightly less complicated. One simultaneously has a dominant
contribution of $s$--wave annihilation into the previous $J^{PC}= 1^{--}$ antisymmetric state, 
$\phi_A$, plus a contamination of $p$--wave annihilation into  $0^{++}$ and $2^{++}$, 
{\it i.e.}, with kaon pairs in the symmetric state $\phi_S = (K_S K_S - K_L K_L)/ \sqrt 2$. 
The coherent addition of these two annihilation amplitudes leads again to a state like 
$\phi_r$, where $r$ now measures the relative strength of the $p$-- to $s$--wave channels. 
Values for this new $r$ of the same order of magnitude as before could be achieved 
in $p\bar{p}$ annihilations at rest by using appropriately 
polarized $\bar{p}$'s and modifying the target densities \cite{CPLEAR2}. Unitary evolution in
free space up to time $T$ leads again to the desired state (\ref{stateN}). In
Ref.~\cite{AlGi} this left--right asymmetric state  has revealed very useful for
Bell--type tests.

As in Ref.~\cite{AlGi}, we consider two mutually exclusive measurements of either strangeness 
or lifetime
to be performed, at will, on each one of the two kaons at a time $T\gsim 10\, \tau_S$, {\it i.e.}, 
when the two kaons are reasonably far away from each other to fulfil the locality requirement 
(for details on how these measurements must be carried out, see Ref.~\cite{AlGi}).
Such interval of $T$ imply a very small value (neglected in what follows) of $|R^\prime|$: 
$|R^\prime|\le 7\cdot 10^{-3}|r|<<1$, and $|R|={\cal O}(1)$.
The following alternative joint measurements will be considered in our argumentation: \\
\{1\} Strangeness on both left and right beams; \\ 
\{2\} Strangeness on the left and lifetime on the right; \\
\{3\} Lifetime on the left and strangeness on the right; \\
\{4\} Lifetime on both left and right beams. \\
Being weak and strong interaction eigenstates related by 
$K_S = (K^0+\bar{K}^0)/\sqrt{2}$ and
$K_L = (K^0-\bar{K}^0)/\sqrt{2}$ ignoring $CP$--violation effects,
state (\ref{stateN}) can be conveniently  rewritten for settings \{1\}, \{2\} and \{3\} as
follows:
\begin{eqnarray}
\label{uno}
\Phi_{\{1\}} &=& {1 \over 2\sqrt{ 2 + |R|^2}} \left[R\, K^0 K^0 +
R\, \bar{K}^0\bar{K}^0 \right. \\
&&\left. +(2-R)\, \bar{K}^0 K^0
-(2+R)\, K^0 \bar{K}^0 \right] , \nonumber \\
\label{due}
\Phi_{\{2\}} &=& {1 \over \sqrt{2(2 + |R|^2)}}
\left[-K^0 K_S + \bar{K}^0 K_S \right. \\
&&\left. +(1+R)\, K^0 K_L +(1-R)\, \bar{K}^0 K_L \right] , \nonumber \\
\label{tre}
\Phi_{\{3\}} &=& {1 \over \sqrt{2(2 + |R|^2)}}
\left[ K_S K^0 - K_S \bar{K}^0 \right. \\
&&\left. -(1-R)\, K_L K^0 -(1+R)\, K_L\bar{K}^0 \right] , \nonumber
\end{eqnarray} 
while, for setting \{4\}, one has $\Phi_{\{4\}} \equiv \Phi$ with $R^\prime =0$.

Now, let us consider the particular case in which $R=-1$ \cite{r-1}. 
We shall refer to the corresponding QM
state as to {\it Hardy's state}. For it, QM predicts \cite{diff-hardy,comment1}: 
\begin{eqnarray}
\label{non-zero}
P_{\rm QM}(K^0,\bar{K}^0)&=&\eta \bar\eta /12 , \\
\label{zero1}  
P_{\rm QM}(K^0,K_L)&=&0 , \\
\label{zero2}  
P_{\rm QM}(K_L,\bar{K}^0)&=&0 , \\
\label{quasi-zero}  
P_{\rm QM}(K_S,K_S)&=& 0 ,
\end{eqnarray}
where $\eta$ and $\bar\eta$ are the ${K}^0$ and $\bar{K}^0$ detection efficiencies of the 
experiment. 
The proof of Bell's theorem without inequalities consists in showing that this set of QM
results is incompatible with LR. This we do by adapting Hardy's argument to our specific case.

It is easy to reproduce the prediction of Eq.~(\ref{non-zero}) under LR. 
Introduce a local hidden--variable model with a normalized probability 
distribution $\rho(\lambda)$ 
($\int_{\Lambda} d\lambda\, \rho(\lambda)=1$) for which, if the pair is created in the
state $\lambda$, with $\lambda \in \Lambda_{0,\bar{0}}$, the single kaon
probabilities to detect a $K^0$ on the left and a $\bar{K}^0$ on the right are 
$0<p_l(K^0|\lambda)\leq 1$ and
$0<p_r(\bar{K}^0|\lambda)\leq 1$,
respectively. These functions and the hidden--variable distribution can be chosen such that:
\begin{eqnarray}
\label{prima}
P_{\rm LR}(K^0,\bar{K}^0)&\equiv&\int_{\Lambda} d\lambda\, \rho(\lambda)\, p_l(K^0|\lambda)\,
p_r(\bar{K}^0|\lambda) = \eta \bar\eta /12 \nonumber \\
&& \leq \int_{\Lambda_{0,\bar{0}}} d\lambda \, \rho(\lambda)\equiv \mu (\Lambda_{0,\bar{0}}) .
\end{eqnarray}

The necessity to reproduce, in LR, predictions (\ref{zero1}) and (\ref{zero2}) has
the following effects. Suppose that in a run of an experiment measuring strangeness
on both sides at proper time $T$, a detection of a $K^0$ on the left and a $\bar{K}^0$ 
on the right occurred. If quantum mechanical prediction (\ref{non-zero}) 
is correct, such an event will be actually observed sometimes.
From the fact that a $K^0$ has been observed on the left for this
specific event, through Eq.~(\ref{zero1}) we can
infer that if lifetime (and not strangeness) had been measured on the right 
with an ideal (efficiency one) detector, one would
have observed a $K_S$. Following the Einstein, Podolsky and Rosen's
condition for the existence of an {\it element of physical reality} (EPR) \cite{epr}, 
the above prediction, made with certainty and without disturbing the right
going particle, permits to assign an EPR to the kaon on the right, the fact 
of being a $K_S$ \cite{gen-epr}:
\begin{equation}
\label{right}
p_r(K_S|\lambda)=1\, , \,\, \forall \, \lambda \in \Lambda_{0,\bar{0}} \, .
\end{equation}
Such an EPR existed independently of any measurement performed on the left going
kaon. In fact, according to the {\it locality assumption},
when the two kaons are space--like separated, the EPR's belonging to 
one kaon cannot be created nor influenced by a measurement made on the other kaon. 
From the fact that a $\bar{K}^0$ has been observed on the right,
by applying a similar argument to the prediction of Eq.~(\ref{zero2}) one 
concludes that the kaon on the left has an EPR, again corresponding to the fact
of being a short living kaon:
\begin{equation}
\label{left}
p_l(K_S|\lambda)=1\, , \,\, \forall \, \lambda \in \Lambda_{0,\bar{0}} \, .
\end{equation}
Imposing locality, the same EPR on the left would have
existed if lifetime (and not strangeness) had been measured on the right.  
For {\it all} the runs of the joint strangeness measurements 
which gave the result $(K^0,\bar{K}^0)$ 
(a fraction $\eta \bar\eta /12$ of the total), we then expect that if one had instead 
measured lifetime along both beams with ideal detectors, one would have
obtained the outcome 
$(K_S,K_S)$. This contradicts QM prediction (\ref{quasi-zero}) since,
through Eqs.~(\ref{prima})--(\ref{left}), it requires:
\begin{eqnarray}
\label{seconda}
P_{\rm LR}(K_S,K_S)&\equiv&\int_{\Lambda} d\lambda\, \rho(\lambda)\, p_l(K_S|\lambda)\,
p_r(K_S|\lambda) \nonumber \\
&\geq &\mu (\Lambda_{0,\bar{0}}) \geq \eta \bar\eta /12 . 
\end{eqnarray}       

To prove whether LR is refuted by Nature, the quantities of Eqs.~(\ref{non-zero})--(\ref{quasi-zero}) 
must then be measured. Such a Hardy--type experiment requires perfect
$K_{L,S}$ detection, but only moderate (strictly, non--vanishing) $K^0$ and
$\bar{K^0}$ detection efficiencies. Let us consider then 
the real experimental possibilities. As discussed in ref.~\cite{AlGi}, 
to measure the strangeness of a neutral kaon in a beam, a piece of ordinary,
nucleonic matter has to be placed at the appropriate (time--of--flight) distance $T$, thus
inducing distinct ${K^0}N$ {\it vs} $\bar{K^0}N$ strangeness--conserving strong reactions which
allow for unambiguous ${K^0}$ {\it vs} $\bar{K^0}$ identification \cite{CPLEAR,ghi}. The situation is
then quite analogous to that encountered in photon polarization measurements, including the 
low detection efficiency effects. Indeed, the need to perform the strangeneness measurement at a given
instant $T$ requires that the piece of matter which induces the kaon--nucleon reaction has to be
rather thin. But then the probability of interaction is considerably reduced \cite{CPLEAR}, at
least for ordinary materials and kaon velocities. However, even if this translates into   
$\eta$, $\bar\eta$ well below 1, our argumentation remains valid. It only requires an ideal efficiency
for the alternative, lifetime measurements at the same instant $T$. In this case, the previous 
piece of material has to be removed in such a way that the neutral kaon continues its
propagation in free space after time $T$. If it is observed to decay shortly after $T$ (mostly
into two easily detectable pions) the neutral kaon has been measured to be a $K_S$ not only at
the decay point but also at the relevant time $T$, since there are no $K_S$--$K_L$ oscillations
in free space. As discussed in refs.~\cite{eber,AlGi}, some $K_S$--$K_L$ misidentifications
will appear (moreover, the two states $K_S$ and $K_L$ are not strictly orthogonal), but only at
an acceptably low level. Therefore, the $K_{L,S}$ detection efficiencies 
seem to be sufficiently close to one.  

\vspace{-0.8cm}
Neutral kaon pairs in the state (\ref{stateN}) seem thus to offer an excellent opportunity to
discriminate between QM and LR in experimental tests quite close to the original proposal by
Hardy. This requires the measurement of the four joint--probabilities 
$P(K^0,\bar{K}^0),  P(K^0,K_L),  P(K_L,\bar{K}^0)$ and $P(K_S,K_S)$, using alternative experimental
setups fulfilling the conventional locality conditions. In order to confirm QM, the latter three
probabilities have to be compatible with zero within experimental errors, {\it i.e.}, no events
should survive after background subtraction \cite{backg} in the three
corresponding runs. Once these null results are (most probably) confirmed one has to look at
events corresponding to strangeness measurements on both sides. The compatibility or not with
zero of the (similarly corrected) number of $(K^0, \bar{K}^0)$ events decides either
against QM or against LR. Needless to say, null measurements cannot be strictly performed but
have allowed us for a conceptually very simple Hardy--type test. More realistic but more involved
treatments would require the use of inequalities, quite in line with the
conventional Bell--type tests and will be discussed elsewhere \cite{long}. 
Our purpose here was another one: to insist on the usefulness of kaon pairs in this kind
of discussions and particularly in those based on Hardy's argument. On the one hand, no
post--selection is required, in contrast to other (photonic) Hardy--experiments. On the other hand, thanks
to the fact that in order to complete Hardy's argument perfect efficient detection is needed for
{\it just one} measurement type (at variance with Bell--type analyses), a
promising possibility to close the efficiency (as well as the locality) loophole has been open. 

This work has been partly supported by the EURODAPHNE EEC--TMR 
program CT98--0169 and by I.N.F.N. Discussions with E. Santos are acknowledged. 

 
\end{multicols}

\begin{references}
\bibitem{r-1} It is possible to obtain $R=-1$
if a beryllium (carbon) regenerator with thickness $d = 2.83$ mm ($d = 0.78$ mm)
is used in addition to a detection time $T=11.09\, \tau_S$ ($T=11.31\, \tau_S$)
and kaon pairs created at a $\phi$--factory ($p\bar{p}$ machine).
\bibitem{diff-hardy} In the standard Hardy--like proofs
\cite{Ha93,boschi,Wu-Kar-Ca},
the probabilities corresponding to our (\ref{zero1}), (\ref{zero2}) and (\ref{quasi-zero}) 
are perfectly vanishing. We have: $P_{\rm QM}(K^0,K_L)/(\eta \eta_\tau)=
P_{\rm QM}(K_L,\bar{K}^0)/(\bar{\eta} \eta_\tau)\simeq (2/3)({\cal{R}}e\, \epsilon)^2 
\simeq 2\cdot 10^{-6}$ (because of $CP$ non--conservation), $\eta_\tau$ being the
efficiency for lifetime measurements, and
$P_{\rm QM}(K_S,K_S)= |R^\prime|^2/3= 8\cdot 10^{-13}$ ($5\cdot 10^{-11}$) for
the experiment at a $\phi$--factory ($p\bar{p}$ machine). 
Nevertheless, this will not prevent us from deriving a contradiction between LR and QM.
\bibitem{comment1}
Note that for $R=1$, a similar contradiction is obtained between LR and the following 
predictions of QM: $P_{\rm QM}(\bar{K}^0,K^0)= \eta \bar\eta /12$,
$P_{\rm QM}(\bar{K}^0,K_L)=
P_{\rm QM}(K_L,K^0)=
P_{\rm QM}(K_S,K_S)=0$.
Unfortunately, a test corresponding to this situation is more difficult to realize:
unless one uses a large detection time $T$, 
with a consequent considerable loss of statistics with respect to the case with
$R=-1$, one cannot prepare an experimental set--up with $R=1$.
$R =\pm 1$ are the only values that allow to prove the contradiction.
\end{references}
\end{document}